\documentclass[titlepage,12pt,a4paper,reqno]{article}
\usepackage{amsmath,amssymb,amsfonts,graphics,setspace,tensor}
\usepackage[linktocpage=true,colorlinks,pdftex]{hyperref}
\usepackage{xcolor}
\colorlet{linkequation}{blue}
\usepackage{bm}
\usepackage{doi}
\usepackage{hyperref}
\hypersetup{colorlinks, citecolor=violet, filecolor=black, linkcolor=black, urlcolor=blue}
\newcommand*{\refeq}[1]{%
  \begingroup
    \hypersetup{
      linkcolor=linkequation,
      linkbordercolor=linkequation,
    }%
    \ref{#1}%
  \endgroup
}
\allowdisplaybreaks

\begin{document}


\begin{titlepage}

\centerline{\Large \bf Finsler Geometries from Topological Electromagnetism} 
\vskip 1cm
\centerline{\bf Adina V. Cri\c{s}an$^{*}$ and Ion V. Vancea$^{\dagger}$
\footnote{\bf I. V. Vancea is the corresponding author. \bf}
}
\vskip 0.5cm
\centerline{\sl $^{*}$Department of Mechanical Systems Engineering}
\centerline{\sl Technical University of Cluj-Napoca}
\centerline{\sl 103 – 105 Muncii Bld., Cluj-Napoca, Romania} 
\centerline{\sl $^{\dagger}$Group of Theoretical Physics and Mathematical Physics,}

\centerline{\sl Department of Physics, Federal Rural University of Rio de Janeiro}
\centerline{\sl Cx. Postal 23851, BR 465 Km 7, 23890-000 Serop\'{e}dica - RJ,
Brazil}
\centerline{
\texttt{\small adina.crisan@mep.utcluj.ro;
ionvancea@ufrrj.br
} 
}
\vspace{0.5cm}

\centerline{15 May 2020}

\vskip 1.4cm
\centerline{\large\bf Abstract}
We analyse the Finsler geometries of the kinematic space of spinless and spinning electrically charged particles in an external Ra\~{n}ada field. We consider the most general actions that are invariant under the Lorentz, electromagnetic gauge and reparametrization transformations. The Finsler geometries form a set parametrized by the gauge fields in each case. We give a simple method to calculate the fundamental objects of the Finsler geometry of the kinematic space of a particle in a generic electromagnetic field. Then we apply this method to calculate the geodesic equations of the spinless and spinning particles. Also, we show that the electromagnetic duality in the Ra\~{n}ada background induces a simple dual map in the set of Finsler geometries. The duality map has a simple interpretation in terms of an electrically charged particle that interacts with the electromagnetic potential and a magnetically charged particle that interacts with the dual magnetoelectric potential. We exemplify the action of the duality map by calculating the dual geodesic equation.

\vskip 0.7cm 

{\bf Keywords}: Finsler geometry; topological electrodynamics; spinning particle; Ra\~{n}ada field.
\noindent

\end{titlepage}


\section{Introduction}

The study of the topological solutions to Maxwell's equations in vacuum, firstly proposed by Trautman and Ra\~{n}ada in \cite{Trautman:1977im,Ranada:1989wc,Ranada:1990}, has revealed so far a rich interplay between physical systems and mathematical structures which was previously unexpected in the realm of classical electrodynamics and classical field theory \cite{Fadeev:1997}. Since then, the 
subject of the topological electromagnetic fields has gain momentum with very interesting problems investigated recently, such as the existence of topological solutions of the Einstein-Maxwell theory \cite{Thompson:2014pta,Kopinski:2017nvp,Vancea:2017tmx,Silva:2018ule} and of the non-linear electrodynamics 
\cite{deKlerk:2017qvq,Goulart:2016orx,Hoyos:2015bxa,Alves:2017zjt,Alves:2017ggb,Nastase:2018gjs}. Also, it has been shown that there are interesting mathematical structures that can be associated to the physical systems with topological electromagnetic fields and play an important role in their dynamics, such as twistors \cite{Dalhuisen:2012zz}, fibrations \cite{Arrayas:2017xvo} and rational functions \cite{Lechtenfeld:2017tif,Kumar:2020xjr}
(see for recent reviews \cite{Arrayas:2017sfq,Vancea:2019zdl,Vancea:2019yjt}). An interesting discussion of the mathematical structures of the axioms of the topological electromagnetism can be found in \cite{Delphenich:2003pz}. Due to the utmost importance of Maxwell's equations, it is essential to further investigate the properties of the topological electromagnetic fields from both physical and mathematical point of view. 

An important mathematical structure that has applications in the geometrization of the electromagnetism, gravity and field theory in general is the Finsler space. In the case of the charged particles in the electromagnetic external field, the Finsler geometry emerges as a natural structure on the kinematic space. The relation between the Finsler geometry and the particle kinematics has been known for some time \cite{Randers:1941gge,Holland:1982bf,Asanov:1985dg,Ingarden:1987,Beil:1987zc,Hsu:1994hw} and it is based on the homogeneity of the Lagrangian functional which allows one to formulate the particle kinematics as a geodesic problem on the Finsler space. In general, the Finsler metric depends on $q/m$ where $q$ is the electric charge of the particle and $m$ is its mass. Due to this dependency, different geometries are generated for different types of particles as well as for different electromagnetic backgrounds. Another interesting feature of the description of the electromagnetism in terms of Finsler spaces is the appearance of alternate gauge conditions for classical charged particles which are due to general relativistic metrics \cite{Beil:1991vv} (see for general references \cite{Rund:1959,Asanov:1986aj,Hehl:1994ue} and \cite{Bao:2020} for a recent pedagogical introduction into the subject).

In this paper, we are going to construct the Finsler geometries 
determined by the topological solutions to Maxwell's equations and calculate the main geometrical objects. Our approach to this problem is different from the literature in that we are considering here the particle action that is invariant, besides the Lorentz and the electromagnetic gauge transformations, under the world-line reparametrization, too. In this formulation, the kinematic space forms a set $\mathfrak{X}[x,\dot{x}; \Phi ]$ parametrized by the world-line einbein $e$ and the electromagnetic potentials $A_{\mu} (x)$. 
We denote these variables collectively by $\Phi$. Each of these spaces has its own Finsler geometry and a particular Finsler geometry can be selected by choosing the gauge fields. The gauge fixing is useful for analysing the physical properties of the system. On the other hand, from the geometrical point of view, it is more interesting to have the geometry defined generally by leaving the gauge fields arbitrary. Thus, the gauge fields can be interpreted as parameters that enter the Lagrangian of the particle as Lagrange multipliers of the first class constraints. In general, the Lagrangians of the relativistic systems have singular Hessian matrices. This property is due to the alternating sign of the pseudo-riemannian metric. However, the physical kinematics of the classical massive particles takes place in the future light-cone defined by the relation $\eta_{\mu \nu} \dot{x}^{\mu} \dot{x}^{\nu} > 0$. Since this relation defines a conical submanifold $\mathcal{M}^{\vee}_{x}(\mathcal{M}) \subset \mathcal{T}_{x}(\mathcal{M})$ one can construct a Finsler space on the physical kinematic space. We use this generic feature of the particle kinematics to formulate the Finsler geometry of $\mathfrak{X}[x,\dot{x}; \Phi ]$ in terms of the topological electromagnetic field for a spin zero  and a spinning relativistic particle \cite{Rivas:1989aw}.

The paper is organized as follows. In Section 2, we give the Finsler geometry of the kinematic space for a relativistic particle in
an arbitrary external electromagnetic field. We use the most general particle action in which \emph{all symmetries} are manifest, i. e. the Lorentz symmetry, electromagnetic gauge and world-line reparametrization \cite{Brink:1976sz,Brink:1976uf,Galvao:1980cu}. The main problem here is to find the fundamental geometrical objects such as the Finsler metric, the inverse metric, the angular metric tensor and the Cartan metric. We give a simple method to find these objects which are in general non-linear and non-polynomial in all variables.
In Section 3, we apply the results from the previous section to determine the Finsler geometry for the particular case of 
an electrically charged particle in a topological solution to Maxwell's equations. These type of solutions, which we call Ra\~{n}ada fields, are parametrized in terms of two complex scalar fields associated to the electric and magnetic field lines, respectively. Here, we calculate the fundamental objects of the Finsler geometries as functions of the complex scalar fields. We find that there is a duality map between the Finsler geometries induced by the electromagnetic duality of Maxwell's equations in vacuum. In Section 4, we formulate the Finsler geometry of the kinematic space of the spinning particle. In this case, the kinematic space is $\mathbb{Z}_2$ graded due to the supersymmetry of the model. However, the Finsler geometry is defined with respect to the commutative variables only. We apply the same method from Section 2 to calculate the geometrical objects of the Finsler geometry. We conclude the paper in the last section. To make this paper self-consistent, we collect the basic concepts of the Finsler geometry that we are using throughout it in the Appendix A. In the Appendix B we give the formulas for the angular metric tensor and the Finsler geodesic equation for the spinless particle. In the Appendix C, we give the same formulas for the spinning particle. Also, we present the geodesic Finsler equation in the dual geometry calculated from the duality map. The main source of Finsler geometry that we use throughout the paper are the references \cite{Rund:1959,Asanov:1986aj,Hehl:1994ue}. We use the natural units $\hbar = c = 1$.

\section{Finsler geometry of kinematic space}

In this section, we formulate the Finsler geometry of the kinematic space 
$\mathfrak{X}[x, \dot{x}; \Phi]$ associated to a particle that moves in an external electromagnetic field. As mentioned in the introduction, the connection between the Finsler geometry and the particle kinematics is known for some time \cite{Randers:1941gge,Holland:1982bf}. The novelty here is that we construct the Finsler geometry for the most general particle Lagrangian that is invariant under all space-time and gauge transformations and give a simple method to calculate the fundamental objects whose definitions and properties can be found in e. g. \cite{Rund:1959, Asanov:1986aj}.  For convenience, we have listed these definitions in the Appendix. 

\subsection{Calculation of the fundamental objects}

The kinematic space of the relativistic particle denoted by $\mathfrak{X}[x, \dot{x}; \Phi]$ is the space of all positions and velocities associated to the particle world-line that, at its turn, is an equivalence class in the set of curves connecting two events from space-time  $x: [ \tau_1 , \tau_2] \to \mathbb{R}^{1,3}$. Here, we designate by $\Phi$ all other variables of the system which are usually associated to the gauge symmetries, the classical spin degrees of freedom and the external fields. In our case, these variables depend only on the world-line coordinate with the only exception of the electromagnetic potentials that depend on position. Some of the parameters from $\Phi$ that are related to the gauge symmetries can be interpreted as Lagrange multipliers for the constraints of the system. In the case of the classical particle, which is our object of study here, the curves $x^{\mu}(\tau)$ are smooth. 

The most general formulation of the spin-$0$ particle in the external electromagnetic potential $A_{\mu}(x)$ is given in terms of an action functional that is manifestly invariant under the Lorentz group $SO(1,3)$, the  $U(1)$ group of the electromagnetic gauge transformations and the world-line reparametrization, and it has the following form \cite{Brink:1976sz,Brink:1976uf}
\begin{equation}
S[x, \dot{x}; e, A] = \int d \tau L[x,\dot{x}; e, A ]
= \int^{\tau_2}_{\tau_1} d \tau 
\left[
\frac{1}{2e} \eta_{\mu \nu} \dot{x}^{\mu} \dot{x}^{\nu}
+ \frac{1}{2} e m^2 + q A_{\mu}(x)\dot{x}^{\mu}
\right]
\, .
\label{action-particle}
\end{equation}
Here, $m$ and $q$ are the mass and electric charge of the particle. 
For the spin-$0$ particle, the set of extra variables is $\Phi = \{e, A \}$, where $e=e(\tau)$ is an auxiliary scalar function (einbein) that makes the reparametrization invariance manifest. By introducing it, the action of the relativistic particle is simultaneously polynomial in the coordinates and non-singular for massless particles. Also, $e$ determines the Hessian matrix of the Lagrangian which is non-null and non-singular for an arbitrary non-vanishing einbein
\begin{equation}
\det 
\left[ 
\frac{\partial^2 L[x, \dot{x}, e ; A ] }{\partial \dot{x}^{\mu} \partial \dot{x}^{\nu}} 
\right] = - \frac{1}{e^4}
\, .
\label{det-L-non-null}
\end{equation} 
Note that, once the equations of motion are imposed on the coordinates and fields, the Hessian given by the equation (\refeq{det-L-non-null}) is formally undetermined since the particles that travel with the speed of light are massless.

For simplicity of notation, let us drop off formulas the arguments of functions unless they are necessary. The dynamics of the particle is obtained by applying the variational principle to the action (\refeq{action-particle}). The equations of motion are
\begin{align}
\frac{d p_{\mu}}{d \tau} - q \partial_{\mu} A_{\nu} \dot{x}^{\nu} & = 0
\, , 
\label{equation-motion-x}
\\
\frac{\dot{x}^{2}}{e^2} - m^2 & = 0 
\, ,
\label{equation-motion-e}
\end{align}
where
\begin{equation}
p_{\mu} := \frac{\partial L}{\partial \dot{x}^\mu}
= \frac{1}{e}\dot{x}_{\mu} + q A_{\mu}
\, .
\label{p-canonical}
\end{equation}
Note that since the einbein is not dynamic, the equation (\refeq{equation-motion-e}) is actually a constraint.

In order to construct the Finsler geometry of the kinematic space, we recall that the points from $\mathbb{R}^{1,3}$ that are causally connected satisfy the relation $\eta_{\mu \nu}\dot{x}^{\mu} \dot{x}^{\nu} > 0$. That shows that the light-cone at $x \in \mathbb{R}^{1,3}$ defines a conical region  of the tangent space at $x$, namely 
\begin{equation}
\mathcal{M}^{\vee}_{x} (\mathbb{R}^{1,3}) = 
\left\{ 
(x, \dot{x} )  \in \mathcal{T}^{*}_{x} (\mathbb{R}^{1,3}) 
\, : \,
\eta_{\mu \nu} \dot{x}^{\mu} \dot{x}^{\nu} > 0
\right\}
\, .
\label{conical-region}
\end{equation} 
The determinant of the Hessian matrix of $L^{2} [x, \dot{x}; e, A ]$ can be written as
\begin{equation}
\det \left[ H_{\mu \nu} \right] = - \left( \frac{2L}{e} \right)^4
\left( 1 + \frac{e}{L} p^2 \right)
\, ,
\quad
p_{\mu}:= 
\frac{\partial L}{\partial \dot{x}^{\mu} }
\, .
\label{Hessian-det-L2}
\end{equation}
From that, we can see that the Hessian is non-singular unless
\begin{equation}
\dot{x}^2 + 3qe A \dot{x} + e^2 q^2 A^2 + \frac{e^2}{2} m^2  = 0
\, .
\label{Hessian-singularity-L2}
\end{equation}
The above equation defines a surface in $\mathcal{T}^{*}_{x} (\mathbb{R}^{1,3}) $ which depends on the world-line parametrization and on the electromagnetic field through $e$ and $A_{\mu}$. However, this surface does not always intersect the set of the causally connected physical events or the world-line. A counterexample is given by the particle in the proper time parametrization and in the absence of the electromagnetic field for which
\begin{equation}
\dot{x}^{2} = - \frac{1}{2} < 0
\, .
\label{hessian-counter-example-singularity-L2}
\end{equation} 
In what follows, we consider only admissible Lagrangians for which the Hessians of $L^{2} [x, \dot{x}; e, A ]$ are non-singular. Then the triplet $\{ \mathbb{R}^{1,3}, \mathcal{M}^{\vee}_{x}(\mathbb{R}^{1,3}), L [x, \dot{x}; e, A ] \}$ defines a four-dimensional Finsler space. This induces a Finsler structure on  $\mathfrak{X}[x,\dot{x}; e,  A]$
which contains points from the tangent space at $x$ together with the values of $e$ and $A$. Thus, 
the geometries on $\mathfrak{X}[x,\dot{x}; e,  A]$ are parametrized by the einbein $e(\tau)$ and the four-vector potential $A_{\mu}(x)$, hence the notation.

The main geometrical objects of the Finsler geometries are defined by the relations reviewed in the Appendix which are proved in e. g. \cite{Rund:1959}. Our main task here is to determine the formulas of the fundamental tensors and the geodesic equations for the geometries of $\mathfrak{X}[x,\dot{x}; e,  A]$. In order to do that, the primary problem to be solved is to invert the Finsler metric. Here, we give a simple procedure to obtain the inverse metric as a non-linear and non-polynomial function on variables and parameters as follows. By definition, the fundamental metric tensor $g_{\mu \nu}[x,\dot{x}; e, A]$ can be written as
\begin{equation}
g_{\mu \nu} = \frac{1}{2} 
\frac{\partial^2 L^{2}}{\partial \dot{x}^{\mu} \partial \dot{x}^{\nu}}
= \frac{\partial L}{\partial \dot{x}^{\mu}}
\frac{\partial L}{\partial \dot{x}^{\nu}}
+ L 
\frac{\partial^2 L}{\partial \dot{x}^{\mu} \partial \dot{x}^{\nu}}
\, .
\label{metric-Finsler-1}
\end{equation}
Then, by using the action (\refeq{action-particle}), we can write the equation (\refeq{metric-Finsler-1}) as follows
\begin{equation}
g_{\mu \nu} = \frac{L}{e} 
\left[
\eta_{\mu \nu} 
+
\frac{e}{L} p_{\mu} p_{\nu}
\right]
\, .
\label{metric-Finsler-2}
\end{equation}
We look for the inverse metric $g^{\mu \nu}$ of the following form
\begin{equation}
g^{\mu \nu} = \frac{e}{L}
\left[
\alpha \eta^{\mu \nu} + \frac{L}{e} \beta 
\left( p^{-1} \right)^{\mu} \left( p^{-1} \right)^{\nu}
\right]
\, ,
\quad
\left( p^{-1} \right)^{\mu} = \frac{p^{\mu}}{p^2}
\, ,
\quad
p^2 = \eta^{\mu \nu} p_{\mu} p_{\nu}
\, ,
\label{metric-inverse-ansatz}
\end{equation}
where $\alpha$ and $\beta$ are real functions. For the inverse metric to exists, it is necessary that
\begin{equation}
p^{2} \neq 0
\, ,
\label{metric-inverse-existence-1}
\end{equation}
where $p^{\mu}$ is given by the equation (\refeq{p-canonical}). From this definition, we can see that the presence of the electromagnetic field lifts in general the singularity in momenta. After some simple algebra, one arrives at the following coefficients $\alpha$ and $\beta$
\begin{equation}
\alpha = 1
\, ,
\quad
\beta = - \frac{e p^2}{L\left(1+ \frac{L}{e p^2}\right)}
\, .
\label{metric-inverse-coefficients}
\end{equation}
Then by plugging the coefficients from the equation (\refeq{metric-inverse-coefficients}) into the inverse metric from the equation (\refeq{metric-inverse-ansatz}) we obtain the following expression
\begin{equation}
g^{\mu \nu} = \frac{e}{L}
\left[
\eta^{\mu \nu}
-\frac{p^{\mu} p^{\nu}}{p^{2} + \frac{L}{e}} 
\right]
\, .
\label{metric-inverse-final-form}
\end{equation}
Again, in order for $g^{\mu \nu}$ to exist, the following condition must be satisfied
\begin{equation}
p^{2} + \frac{L}{e} \neq 0 
\, .
\label{metric-inverse-existence-2}
\end{equation}
This condition is in general valid. Next, we calculate the Finsler-Christoffel symbols by plugging the equations (\refeq{metric-inverse-ansatz}) and (\refeq{metric-inverse-final-form})  into the definition (\refeq{Finsler-Christoffel-symbols-definition}). The results is the following formula
\begin{align}
{\Gamma^{\mu}}_{\nu \rho} & =
\frac{e}{2L} 
\left( \eta^{\mu \lambda} 
- \frac{p^{\mu} p^{\lambda}}{p^2 + \frac{L}{e}}
\right)
\left[
\frac{1}{e}
	\left(
	\eta_{\nu \lambda} \partial_{\rho} L +
	\eta_{\rho \lambda} \partial_{\nu} L -
	\eta_{\nu \rho} \partial_{\lambda} L
	\right)
\right.
\nonumber
\\
& +
\left.
	\partial_{\rho} \left( p_{\nu} p_{\lambda} \right) +
	\partial_{\nu} \left( p_{\rho} p_{\lambda} \right) -
	\partial_{\lambda} \left( p_{\nu} p_{\rho} \right)
\right]
\, . 
\label{Finsler-Christoffel-symbols-charged-particle}
\end{align}
With these relations at hand, we can write down the Finsler geodesic equation defined by the equation (\refeq{geodesic-equation-Finsler-definition}). Before presenting it, it is useful to introduce the following shorthand notations in which the equations have a more compact form
\begin{equation}
\boldsymbol{\partial} = \dot{x}^{\mu} \partial_{\mu}
\, ,
\quad
\mathbf{V} = \dot{x}^{\mu} V_{\mu}
\, ,
\label{notations-Finsler}
\end{equation}
where $V_{\mu} (x)$ is an arbitrary four-vector. Also, we recall that the variables $x^{\mu}$ and $\dot{x}^{\mu}$ are independent on each other and that the indices are raised and lowered with the inverse and the direct Minkowski metric, i. e. $V^{\mu} = \eta^{\mu \nu} V_{\nu}$, etc. With these notations, the Finsler geodesic equation takes the following form
\begin{equation}
\frac{d^2 {x}^{\mu}}{d \tau^2 }  +
\frac{e}{2L} 
\left( \eta^{\mu \nu} - \frac{p^{\mu} p^{\nu}}{p^2 + \frac{L}{e}}
\right)
\left[
\frac{1}{e}
	\left( 
	2 x_{\nu} \boldsymbol{\partial} L -
	\dot{x}^{2} \partial_{\nu} L
	\right)
	+ \boldsymbol{\partial} \left( \mathbf{p} p_{\nu} \right)
	- \partial_{\nu} \left( \mathbf{p}^2 \right)
\right]
= \dot{x}^{\mu} \frac{d}{d \tau} \left[ \ln(L)\right]
\, .
\label{geodesic-equation-Finsler-final}
\end{equation} 
For completeness, we give here the angular metric tensor and the Cartan tensor. 
From the definition (\refeq{metric-unit-angular-Finsler-def}) we can immediately write
\begin{equation}
h_{\mu \nu} = \frac{L}{e}
\left[
\eta_{\mu \nu} + \frac{e}{L} p_{\mu} p_{\nu}
\right]
-
\frac{\dot{x}_{\mu} \dot{x}_{\nu}}{L^2}
\, .
\label{metric-angular-spin-zero-particle}
\end{equation}
Also, from the definition (\refeq{Cartan-tensor}) we obtain the following result
\begin{equation}
C_{\mu \nu \rho} = 
\frac{1}{2} \eta_{\mu \nu} 
\frac{\partial L}{\partial \dot{x}^{\rho}}
+
\frac{1}{e} \delta_{\rho (\mu} p_{\nu )}
\, .
\label{Cartan-tensor-Finsler}
\end{equation}
We remark that the fundamental objects of the Finsler geometry obtained above are in general non-linear and non-polynomial in the variables $x^{\mu}$ and $\dot{x}^{\mu}$. This is due to the fact that the inverse metric contains negative powers of these variables.

The fundamental tensors and the Finsler geodesic equation are parametrized by $e$ and $A_{\mu}$ (the potential vector is implicit in $L$). Therefore, to each set of functions $e$ and $A_{\mu}$ corresponds a geometry. This observation is important if we recall that by choosing particular values for $e$ and $A_{\mu}$ one fixes the reparametrization and the electromagnetic gauge symmetries, respectively.
In the classical theory, $e$ can have certain values even if the reparametrization symmetry does not allow these values \cite{Galvao:1980cu}. However, in the quantum theory the gauge structure  must be respected.

\subsection{Action of gauge symmetries on Finsler geometries}

The gauge transformations of the action from the equation (\refeq{action-particle})  induce transformations among the Finsler geometries in the parameters $e$ and $A$. A finite world-line reparametrization is defined by the following transformations 
\begin{equation}
\tau^{\prime} = f(\tau)
\, , 
\quad 
x^{\prime \mu} (\tau) = x^{\mu}( f(\tau) )
\, ,
\quad
e d \tau= e^{\prime} df(\tau)
\, , 
\label{reparametrization-transformation-spin-0-particle}
\end{equation}
where $f$ is an arbitrary smooth function that has the following properties
\begin{equation}
\frac{d f (\tau )}{d \tau} > 0
\, ,
\quad
f(\tau_1) = 0
\, ,
\quad
f(\tau_2 ) = 1
\, .
\label{reparametrization-f-function}
\end{equation}
One can easily verify that 
the fundamental metric tensor, the angular metric tensor and the Cartan tensor
transform under the transformations (\refeq{reparametrization-f-function}) as follows
\begin{equation}
g^{\prime}_{\mu \nu} = g_{\mu \nu}
\, ,
\quad
h^{\prime}_{\mu \nu} = h_{\mu \nu}
\, ,
\quad
C^{\prime}_{\mu \nu \rho} =  \frac{df}{d\tau} C_{\mu \nu \rho}
\, .
\label{reparametrization-of-metric-angular-Cartan}
\end{equation}
The Finsler geodesic equation $F^{\mu} = 0$ is rescaled as
\begin{equation}
F^{\mu} = \left( \frac{df}{d\tau} \right)^{2} F^{\prime \mu} = 0
\, ,
\end{equation} 
which for arbitrary $f$ implies that $F^{\prime \mu} = 0$. 
The infinitesimal reparametrizations form an infinite dimensional continuous group that can be interpreted as a gauge group. The corresponding transformations can be obtained from the finite ones by setting $f(\tau) = 1 + \epsilon(\tau)$, $\vert \epsilon (\tau ) \vert \ll 1$ for all $\tau \in [ \tau_1 , \tau_2 ]$. 

The  action (\refeq{action-particle}) is invariant under the electromagnetic gauge transformations of the following form
\begin{equation}
A^{\prime}_{\mu}(x) = A_{\mu} (x) + \partial_{\mu} \Lambda (x)
\, ,
\label{gauge-transformations-A}
\end{equation}
where $\Lambda (x)$ is a smooth but arbitrary function on $x^{\mu}$'s only. The Lagrangian is invariant under the transformations (\refeq{gauge-transformations-A}) up to a surface term 
\begin{equation}
L^{\prime} = L 
+ q \boldsymbol{\partial} \Lambda
\, .
\label{gauge-transformations-Lagrangian}
\end{equation}
One can easily verify that the Finsler metric tensor,
angular metric tensor and Cartan tensor transform as follows under the gauge transformations
\begin{align}
g^{\prime}_{\mu \nu}  & = 
g_{\mu \nu} + 
q^2 
\left[ 
2 A_{(\mu} \partial_{\nu)} \Lambda +
\partial_{\mu} \Lambda \partial_{\nu} \Lambda
\right]
+ \frac{q}{e}
\left[
\eta_{\mu \nu} \partial_{\rho} \Lambda +
2 \eta_{\rho (\mu} \partial_{\nu)} \Lambda
\right] \dot{x}^{\rho}
\, ,
\label{metric-gauge-transformations-Finsler}
\\
h^{\prime}_{\mu \nu}  & = 
h_{\mu \nu} +
q^2 
\left[
2A_{(\mu}\partial_{\nu)} \Lambda + \partial_{\mu} \Lambda 
\partial_{\nu} \Lambda 
\right] 
+ \frac{q}{e}
\left[
\eta_{\mu \nu} \partial_{\rho} \Lambda +
2 \eta_{\rho (\mu} \partial_{\nu)} \Lambda
\right] \dot{x}^{\rho}
\nonumber
\\
& - \frac{\dot{x}_{\mu} \dot{x}_{\nu}}{L^2}
\sum_{n=1}^{\infty} 
\left( 
- \frac{q}{L} \boldsymbol{\partial} \Lambda 
\right)^{n}
\, ,
\label{angular-metric-gauge-transformations-Finsler}
\\
C^{\prime}_{\mu \nu \rho} & = C_{\mu \nu \rho} +
\frac{q}{2e} \eta_{\mu \nu} \partial_{\rho} \Lambda
+\frac{q}{e} \eta_{\rho (\mu} \partial_{\nu )} \Lambda
\label{Cartan-tensor-gauge-transformations-Finsler}
\, .
\end{align}
The above transformations establish a relationship among different Finsler geometries that can be associated to the same physical system. 
As in the case of the world-line reparametrization, by fixing a gauge one fixes the Finsler geometry on the kinematic space. This is important if one wishes to quantize the system by using the Finsler space structure.

We recall that the gauge symmetry leaves the electromagnetic field invariant, thus inducing an equivalence relation among electromagnetic potentials on physical grounds. By using the same argument, we can define and equivalence relation among the Finsler geometries if the fundamental tensor objects are related as in the equations (\refeq{metric-gauge-transformations-Finsler}) - (\refeq{angular-metric-gauge-transformations-Finsler}). The practical application of this definition is that it guarantees the equivalence of the classical physics among all Finsler geometries connected by gauge transformations.

\subsection{Dual Finsler geometries from Ra\~{n}ada fields }

In this section, we will apply the results obtained previously to the relativistic particle that moves in a topological electromagnetic field which is a solution to Maxwell's equations in vacuum. The most general form of a topological electromagnetic field was given by Ra\~{n}ada in \cite{Ranada:1989wc,Ranada:1990} and in what follows we will focus on the Ra\~{n}ada fields\footnote{The physics of a relativistic particle moving in a particular type of topological electromagnetic field was discussed previously in \cite{Arrayas:2010xi} but without any reference to the associated Finsler space.}. 

The dynamics of the electromagnetic field in vacuum is given by the well-known homogeneous Maxwell equations
\begin{align}
\partial_{\mu} F^{\mu \nu} & = 0
\, ,
\label{Gauss-Ampere}
\\
\epsilon^{\mu \nu \rho \sigma} \partial_{\nu} 
F_{\rho \sigma} 
& = 0
\, ,
\label{Gauss-Faraday}
\end{align}
where $F_{\mu \nu} = \partial_{\mu } A_{\nu} - \partial_{\nu } A_{\mu}$.
The most general topological solutions of Maxwell's equations 
(\refeq{Gauss-Ampere}) and (\refeq{Gauss-Faraday})
are given by the 
Ra\~{n}ada fields which are parametrized by 
two smooth complex scalar fields $\phi: \mathbb{R}^{1,3} \to \mathbb{C}$ and $\theta:\mathbb{R}^{1,3} \to \mathbb{C}$ and have the following form \cite{Ranada:1989wc,Ranada:1990}
\begin{align}
F_{\mu \nu} & = \frac{\sqrt{a}}{2 \pi i}
\frac{\partial_{\mu} \bar{\phi} \partial_{\nu} \phi
- \partial_{\nu} \bar{\phi}\partial_{\mu} \phi}{(1+\vert \phi \vert^2)^2}
\, ,
\label{Ranada-F-field}
\\
\tensor[^{*}]{F}{_{\mu}_{\nu}} & = \frac{\sqrt{a}}{2 \pi i}
\frac{\partial_{\mu} \theta \partial_{\nu} \bar{\theta}
- \partial_{\nu} \theta \partial_{\mu} \bar{\theta}}{(1+\vert \theta \vert^2)^2}
\, , 
\label{Ranada-G-field}
\end{align}
where $a$ is a positive real constant and $\tensor[^{*}]{F}{_{\mu}_{\nu}} = \frac{1}{2}\epsilon_{\mu \nu \rho \sigma}F^{\rho \sigma}$. The fields $\phi$ and $\theta$ describe the electric and magnetic lines through the equations $\phi(x) = \mbox{constant}$ and $\theta(x) = \mbox{constant}$, respectively. Thus, the Ra\~{n}ada fields (\refeq{Ranada-F-field}) and (\refeq{Ranada-G-field}) are adequate to describe the topological and geometrical properties of the electromagnetic field in terms of its integral lines.

Let us determine the Finsler structures of the kinematic space of the charged particle in the Ra\~{n}ada field given by the equations (\refeq{Ranada-F-field}) and (\refeq{Ranada-G-field}) above. 
The electromagnetic potentials $A_{\mu}$ and $C_{\mu}$ of the fields $F_{\mu \nu}$ and $\tensor[^{*}]{F}{_{\mu}_{\nu}}$ are given by the following relations
\begin{align}
A_{\mu} & = \frac{\sqrt{a}}{4 \pi i}
\left(
\frac{\bar{\phi} \partial_{\mu} \phi - \phi \partial_{\mu} \bar{\phi}}{1+ \vert \phi \vert^{2}}
\right)
\, ,
\label{A-potentials}
\\
C_{\mu} & = \frac{\sqrt{a}}{4 \pi i}
\left(
\frac{\bar{\theta} \partial_{\mu} \theta - \theta \partial_{\mu} \bar{\theta}}{1+ \vert \theta \vert^{2}}
\right)
\, .
\label{C-potentials}
\end{align}
Since Maxwell's equations in vacuum are invariant under the electromagnetic duality transformations, the scalars $\phi$ and $\theta$ are not independent on each other. The constraints imposed by the duality on $\phi$ and $\theta$ are given by the following non-linear equations \cite{Ranada:1989wc,Ranada:1990}
\begin{align}
\frac{1}{\left( 1+ \vert \phi \vert^2\right)^2}
\epsilon_{ijk} \partial_i \phi \partial_j \bar{\phi}
& =
\frac{1}{\left( 1+ \vert \theta \vert^2\right)^2}
\left( 
\partial^{0}\bar{\theta} \partial_{k} \theta
-
\partial^{0}\theta \partial_{k} \bar{\theta}
\right)
\, ,
\label{duality-phi-theta}
\\
\frac{1}{\left( 1+ \vert \theta\vert^2 \right)^2}
\epsilon_{ijk} \partial_i \bar{\theta} \partial_j \theta
& =
\frac{1}{\left( 1+ \vert \phi \vert^2\right)^2}
\left( 
\partial^{0}\bar{\phi} \partial_{k} \phi
-
\partial^{0}\phi \partial_{k} \bar{\phi}
\right)
\, .
\label{duality-theta-phi}
\end{align}
The equations (\refeq{A-potentials}) and (\refeq{C-potentials}) show that we can define two Finsler metric structures with the fundamental metric tensors $g_{\mu \nu} [x, \dot{x}, e; A ]$ and $\tilde{g}_{\mu \nu} [y, \dot{y}; f, C ]$ associated to the Finsler spaces $\mathfrak{X}[x,\dot{x}; e , A]$ and $\mathfrak{Y}[y,\dot{y}; f , C]$. We call these \emph{dual Finsler geometries} since the scalar fields $\phi$ and $\theta$ are dual to each other according to the equations (\refeq{duality-phi-theta}) and (\refeq{duality-theta-phi}). 

An important problem about the dual Finsler geometries given above is whether there is any relationship between them. Instead of addressing this problem by tempting to find a solution to the equations (\refeq{duality-phi-theta}) and (\refeq{duality-theta-phi}) which are non-linear, we are going to solve it by noting that while the space  $\mathfrak{X}[x,\dot{x}; e , A]$ is again the kinematic space of the electrically charged particle that interacts with the topological field $F_{\mu \nu}$, the interpretation of the second space $\mathfrak{Y}[y,\dot{y}; f , C]$ is not so straightforward. If we stick to the representation of the Finsler geometry as a structure of a kinematic space, then the most natural interpretation that can be given to $\mathfrak{Y}[y,\dot{y}; f, C]$ is in terms of a charged particle that would move itself in the dual field $^{*}F_{\mu \nu}$ according to the formal action
\begin{equation}
S[y; f, C] = \int d \tau \tilde{L}[y,\dot{y}; f , C ]
= \int^{\tau_2}_{\tau_1} d \tau 
\left[
\frac{1}{2f} \eta_{\mu \nu} \dot{y}^{\mu} \dot{y}^{\nu}
+ \frac{1}{2} f M^2 + k C_{\mu}(x)\dot{y}^{\mu}
\right]
\, ,
\label{action-particle-magnetic}
\end{equation}
where by analogy to the action from the equation (\refeq{action-particle}), $M$ is the mass of the particle and $k$ is its charge. In this interpretation, it is tempting to formally identify the $(M,k)$ dual particle with the magnetic monopole of magnetic charge $k$ and $C_{\mu}(x)$ with an \emph{external} magnetoelectric potential. Indeed, the magnetic monopole cannot be the source of the Ra\~{n}ada field as it would introduce a Dirac string related to the singularity of $A_{\mu}(x)$ that would be the result of the violation of the corresponding Bianchi identity.  

By using the equations (\refeq{A-potentials}) and (\refeq{C-potentials}) into the equation (\refeq{metric-Finsler-2}), we can write down immediately the fundamental metric tensors of $\mathfrak{X}[x,\dot{x}; e , A]$ and $\mathfrak{Y}[y,\dot{y}; f , C]$ and we obtain the following relations
\begin{align}
g_{\mu \nu} [x, \dot{x}; e, A ] & = 
\frac{m^2}{2 } \eta_{\mu \nu} -   
\frac{q^2 a}{ 16 \pi^2}
\left(
\frac{\bar{\phi} \partial_{\mu} \phi - \phi \partial_{\mu} \bar{\phi}}{1+ \vert \phi \vert^{2}}
\right)
\left(
\frac{\bar{\phi} \partial_{\nu} \phi - \phi \partial_{\nu} \bar{\phi}}{1+ \vert \phi \vert^{2}}
\right)
\nonumber
\\
& + \frac{q}{e}
\left[
\frac{
\eta_{\mu \nu} 
\left(
\bar{\phi} \partial_{\rho} \phi - \phi \partial_{\rho} \bar{\phi}
\right)
+ 2\eta_{\rho (\mu} 
\left(
\bar{\phi} \partial_{\nu)} \phi - \phi \partial_{\nu)} \bar{\phi}
\right) 
}{1+ \vert \phi \vert^{2}}
\right] \dot{x}^{\rho}
\nonumber
\\
& 
+ \frac{1}{2 e^2}
\left[
\eta_{\mu \nu} \dot{x}^2 + 2 \eta_{\mu \rho} \eta_{\nu \sigma}
\dot{x}^{\rho} \dot{x}^{\sigma}
\right]
\, ,
\label{metric-tensor-A}
\\
\tilde{g}_{\mu \nu} [y, \dot{y}; f, C ] & = 
\frac{M^2}{2 } \eta_{\mu \nu} -   
\frac{k^2 a}{ 16 \pi^2}
\left(
\frac{\bar{\theta} \partial_{\mu} \theta - \theta \partial_{\mu} \bar{\theta}}{1+ \vert \theta \vert^{2}}
\right)
\left(
\frac{\bar{\theta} \partial_{\nu} \theta - \theta \partial_{\nu} \bar{\theta}}{1+ \vert \theta \vert^{2}}
\right)
\nonumber
\\
& + \frac{k}{f}
\left[
\frac{
\eta_{\mu \nu} 
\left(
\bar{\theta} \partial_{\rho} \theta - \theta \partial_{\rho} \bar{\theta}
\right)
+ 2\eta_{\rho (\mu} 
\left(
\bar{\theta} \partial_{\nu)} \theta - \theta \partial_{\nu)} \bar{\theta}
\right) 
}{1+ \vert \theta \vert^{2}} 
\right]
\dot{y}^{\rho}
\nonumber
\\
& 
+ \frac{1}{2 f^2}
\left[
\eta_{\mu \nu} \dot{y}^2 + 2 \eta_{\mu \rho} \eta_{\nu \sigma}
\dot{y}^{\rho} \dot{y}^{\sigma}
\right]
\, .
\label{metric-tensor-C}
\end{align}
The introduction of the formal kinematic space for the dual particle
exploits the similarity between the solutions (\refeq{Ranada-F-field}) and (\refeq{Ranada-G-field}). Beside that, it allows one to establish  the sought for map between the dual Finsler geometries by observing that the fundamental metric tensor $g_{\mu \nu} [x, \dot{x}; e, A ]$ is mapped into $\tilde{g}_{\mu \nu} [y, \dot{y}; f, C ]$ by the following mapping
\begin{equation}
g_{\mu \nu} [x, \dot{x}, e; A ] \to \tilde{g}_{\mu \nu} [y, \dot{y}, f; C ]
\, := \,
\{ x \to y \, , e \to f \, , 
\phi \to - \theta \, , 
m \to M \, , q \to - k 
\}
\, .    
\label{g-tilde-g-map}
\end{equation}
The above relation represents the \emph{duality mapping} of the dual Finsler geometries and it is obviously invertible. It can be applied to derive other fundamental objects on $\mathfrak{X}[x,\dot{x}; e , A]$ and $\mathfrak{Y}[y,\dot{y}; f , C]$ from each other. Its simplicity is due to the introduction of the formal action (\refeq{action-particle-magnetic}) which provides the Finsler metric function $\tilde{L}[y,\dot{y}; f , C ]$
and allows one to interpret the second Finsler space as a formal kinematic space for the movement of a magnetically charged dual particle in the dual field.
The Lagrangians  $L[x,\dot{x}; e , A ]$ and $\tilde{L}[y,\dot{y}; f , C ]$ can also be mapped into each other by the relations (\refeq{g-tilde-g-map}).  
If one does not use the dual particle interpretation of the second Finsler geometry, it is non-trivial to determine a relationship between the dual Finsler geometries corresponding to the Ra\~{n}ada fields since the equations (\refeq{duality-phi-theta}) and (\refeq{duality-theta-phi}) that relate the complex fields to each other are non-linear. We note that there is a relationship between the duality map (\refeq{g-tilde-g-map}) and the electromagnetic duality of Maxwell's equations with sources which end up producing the same terms in the Lagrangians. However, the magnetically charged particle described by 
$\tilde{L}[y,\dot{y}, f ; C ]$ should be formal rather than physical since it must preserve the Bianchi identities for the electromagnetic and magnetoelectric potentials, respectively, which are key to constructing the Ra\~{n}ada fields. 

To illustrate the importance of the knowledge of the duality map between the dual Finsler geometries, we consider the Finsler geodesic equation in $\mathfrak{X}[x,\dot{x}; e , A]$ which is obtained by using the 
Ra\~{n}ada potentials from the equations (\refeq{A-potentials}) and (\refeq{geodesic-equation-Finsler-final}). The results is a non-linear equation that involves non-polynomial terms. The geodesic equation on the dual space $\mathfrak{Y}[y,\dot{y}; f , C]$ can be obtained either by direct calculations or by applying the duality map (\refeq{g-tilde-g-map}) to the geodesic equation on $\mathfrak{X}[x,\dot{x}; e , A]$. Since the results are given by large formulas, they are presented in the Appendix B.

For completeness, we give here the Cartan tensor in the field $A_{\mu}$
\begin{align}
C_{\mu \nu \rho} [x, \dot{x}, e; A ] & = 
+ \frac{q}{e}
\left[
\frac{
\eta_{\mu \nu} 
\left(
\bar{\phi} \partial_{\rho} \phi - \phi \partial_{\rho} \bar{\phi}
\right)
+ 2\eta_{\rho (\mu} 
\left(
\bar{\phi} \partial_{\nu)} \phi - \phi \partial_{\nu)} \bar{\phi}
\right) 
}{1+ \vert \phi \vert^{2}}
\right]
\nonumber
\\
& 
+ \frac{1}{e^2}
\left[
\eta_{\mu \nu} \eta_{\rho \sigma} 
+ 
\eta_{\mu \rho} \eta_{\nu \sigma}
+
\eta_{\mu \sigma} \eta_{\nu \rho}
\right] \dot{x}^{\sigma}
\, ,
\label{Cartan-tensor-Ranada}
\end{align}
The angular metric tensor has a more extended formula which is presented in the Appendix B. The duality map (\refeq{g-tilde-g-map}) provides the corresponding tensor objects in the $C_{\mu}$ field. Since its application is very simple, we leave it as an exercise to the reader.

\section{Finsler geometries from spin-$\frac{1}{2}$ particle in Ra\~{n}ada background}

In this section, we will generalized the construction presented above to the classical spinning particle of one-half spin. The kinematic space of the free spinning particle $\mathfrak{X}_{(\frac{1}{2})}[ x, \dot{x}, \psi, \dot{\psi}; e, \chi , A]$ is constructed over the functions $\{x^{\mu}, \dot{x}^{\mu}, \psi^{\mu}, \dot{\psi}^{\mu}, \psi_5, e, \chi \} $ on the particle world-line. The variables $\{ x^{\mu} \}$ are bosonic and $\{ \psi^{\mu}, \psi_5 \}$ are fermionic. The variables $\{ \psi^{\mu}, \psi_5 \}$ are introduced to define the spin in the classical model. By definition, they satisfy the following Clifford algebra relations
\begin{equation}
\left[ \psi^{\mu} , \psi^{\nu} \right]_{+} = \eta^{\mu \nu}
\, ,
\qquad
\left[ \psi_5 , \psi_5 \right]_{+} = - 1
\, , 
\qquad
\left[ \psi^{\mu} , \psi_{5} \right]_{+} = 0
\, ,
\label{definition-spin-variables}
\end{equation}
where the brackets are anti-commutators.
The most general action of the free spinning particle 
that is invariant under the supersymmetry and reparametrization transformation is obtained by integrating the following Lagrangian \cite{Brink:1976uf}
\begin{equation} 
L_{0} [ x, \dot{x}, \psi, \dot{\psi} ,e, \chi ; A]  =
\frac{1}{2e} \eta_{\mu \nu} \dot{x}^{\mu} \dot{x}^{\nu} + \frac{1}{2} e m^2
+ \frac{i}{2} 
 \left(
\dot{\psi}_{\mu} \psi^{\mu} - \dot{\psi}_{5} \psi_{5}
\right)
- \frac{i}{2e} \chi  \psi_{\mu} \dot{x}^{\mu}
+ im \chi \psi_{5}
\nonumber
\,  .
\label{Lagrangian-free-spinning-particle}
\end{equation}
The functions $e$ and $\chi$ are Lagrange multipliers of the two first class constraints that must be introduced into the action. The electromagnetic potential $A_{\mu} (x)$ plays the role of an external field as in the case of the spin-$0$ particle. The interaction between the spinning particle and the electromagnetic field is expressed by the following Lagrangian
\begin{equation}
L_{int} [ x, \dot{x}, \psi, \dot{\psi}, e, \chi ; A]
= \\
 q  A_{\mu} \dot{x}^{\mu} + \frac{iq}{2} e F_{\mu \nu} \psi^{\mu} \psi^{\nu} 
 \, .
\label{Lagrangian-interaction-spinning-particle}
\end{equation}
Then the total Lagrangian of the spin-$\frac{1}{2}$ particle in the electromagnetic field has the form
\begin{align} 
L_{(\frac{1}{2})}[ x, \dot{x}, \psi, \dot{\psi} ,e, \chi ; A]
& =
\frac{1}{2e} \eta_{\mu \nu} \dot{x}^{\mu} \dot{x}^{\nu} + \frac{1}{2} e m^2
+ \frac{i}{2} 
\left(
\dot{\psi}_{\mu} \psi^{\mu} - \dot{\psi}_{5} \psi_{5}
\right)
- \frac{i}{2e} \chi  \psi_{\mu} \dot{x}^{\mu}
+ im \chi \psi_{5}
\nonumber
\\
& +
q  A_{\mu} \dot{x}^{\mu} + \frac{iq}{2} eF_{\mu \nu} \psi^{\mu} \psi^{\nu} 
\, .
\label{Lagrangian-total-spinning-particle}
\end{align}
We note that $L_{(\frac{1}{2})}[ x, \dot{x}, \psi, \dot{\psi} ,e, \chi ; A]$ is invariant under the super-reparametrization and electromagnetic gauge transformations (see for details \cite{Brink:1976uf}). 

The kinematic space $\mathfrak{X}_{(\frac{1}{2})}[ x, \dot{x}, \psi, \dot{\psi}, e, \chi ; A]$ is $\mathbb{Z}_2$-graded as it contains both bosonic and fermionic variables
\begin{equation}
\mathfrak{X}_{(\frac{1}{2})}[ x, \dot{x}, \psi, \dot{\psi}; e, \chi , A] =
\mathfrak{X}_{0}[ x, \dot{x}; e, A]
\oplus \mathfrak{X}_{1}[ \psi, \dot{\psi}; \chi]
\, .
\label{kinematic-space-spinning-particle}
\end{equation}
The space $\mathfrak{X}_{0}[ x, \dot{x}; e, A]$ is associated to the particle world-line as in the spin-$0$ case and one can try to give it a Finsler space structure in terms of the bosonic coordinates. However, 
we expect that the corresponding Finsler fundamental tensor depend on the anti-commutative variables as well due to the interaction with the electromagnetic field. In what follows we will drop off the arguments of the functions in formulas. The determinant of the Hessian matrix of $L_{(\frac{1}{2})}$ with respect to the bosonic variables $\dot{x}^{\mu}$ is
\begin{equation}
\det 
\left[ 
\frac{\partial^2 L_{(\frac{1}{2})}}{\partial \dot{x}^{\mu} \partial \dot{x}^{\nu} }
\right]
= - \frac{1}{e^4}
\, .
\label{Hessian-spinning-particle}
\end{equation}
As in the case of the spinning particle, the Hessian of $L^{2}_{(\frac{1}{2})}$ is zero only if
\begin{align}
& \dot{x}^2 + 9qe A \dot{x} + 3 e^2 q^2 A^2 + \frac{3}{2} m^2 e^2 
- 3 i e \chi
\left(
\psi \dot{x} + q e A \psi - m \psi_5  
\right) 
\nonumber
\\
& 
+ \frac{3i}{2} e 
\left(
\dot{\psi} \psi - \dot{\psi}_{5} \psi_{5}
\right)
+ \frac{3q}{2} e^2 F S
= 0
\, ,
\label{Hessian-singularity-L2-spinning}
\end{align}
where we have suppressed the summation indices. In the absence of the external field, the real part of equation (\refeq{Hessian-singularity-L2-spinning}) does respect the classical causality.

We define the Finsler space of the spinning particle as in the spin-$0$ case by the triple $\{ \mathbb{S}^{1,3}, \mathcal{M}^{\vee}_{x} (\mathbb{S}^{1,3}), L_{(\frac{1}{2})} \}$ where $\mathbb{S}^{1,3}$ is the superspace associated to the $\mathbb{R}^{1,3}$ such that the coordinates of each point $P \in \mathbb{S}^{1,3}$ are $\{ x^{\mu}, \psi^{\mu} \}$.
The conical region $\mathcal{M}^{\vee}_{x} (\mathbb{S}^{1,3})$ is defined with respect to the bosonic coordinates by the Minkowski metric. The induced Finsler structure on the kinematic space $\mathfrak{X}_{(\frac{1}{2})}[ x, \dot{x}, \psi, \dot{\psi}; e, \chi , A]$ has the same form as the general kinematic space $\mathfrak{X}[ x, \dot{x}; \Phi ]$, therefore we can apply the same method to calculate the fundamental objects.

The Finsler metric is defined with respect to the bosonic variables $\dot{x}^{\mu}$ and it has the following form
\begin{align} 
g^{(\frac{1}{2})}_{\mu \nu}  & = 
\frac{m^2}{2 } \eta_{\mu \nu} -   
\frac{q^2 a}{ 16 \pi^2}
\left(
\frac{\bar{\phi} \partial_{\mu} \phi - \phi \partial_{\mu} \bar{\phi}}{1+ \vert \phi \vert^{2}}
\right)
\left(
\frac{\bar{\phi} \partial_{\nu} \phi - \phi \partial_{\nu} \bar{\phi}}{1+ \vert \phi \vert^{2}}
\right)
\nonumber
\\
& 
+ \frac{q}{e}
\left[
\frac{
\eta_{\mu \nu} 
\left(
\bar{\phi} \partial_{\rho} \phi - \phi \partial_{\rho} \bar{\phi}
\right)
+ 2\eta_{\rho (\mu} 
\left(
\bar{\phi} \partial_{\nu)} \phi - \phi \partial_{\nu)} \bar{\phi}
\right) 
}{1+ \vert \phi \vert^{2}}
\right] \dot{x}^{\rho}
\nonumber
\\
& 
+ \frac{1}{2 e^2}
\left[
\eta_{\mu \nu} \dot{x}^2 + 2 \eta_{\mu \rho} \eta_{\nu \sigma}
\dot{x}^{\rho} \dot{x}^{\sigma}
\right]
+ \frac{i}{2 e} \eta_{\mu \nu}
\left(
\dot{\psi}_{\rho} \psi^{\rho} - \dot{\psi}_5 \psi_5
\right)
\nonumber
\\
&
-\frac{i}{2e} \chi
\left[
\frac{\sqrt{a}\, e}{2\pi i}
\left(
\frac{\psi_{(\mu} \bar{\phi} \partial_{\nu)} \phi - 
	\psi_{(\mu}\phi \partial_{\nu)} \bar{\phi}}
	{1+ \vert \phi \vert^{2}}
\right) 
+ 
\eta_{\mu \nu }
\left(
\psi_{\rho} \dot{x}^{\rho}
- 2  m \psi_5  
\right)
\right]
\nonumber
\\
& +
\eta_{\mu \nu}
\frac{q \sqrt{a}}{8\pi i}
\left(
\frac{\partial_{\rho} \bar{\phi} \partial_{\sigma} \phi - 
	\partial_{\rho}\phi \partial_{\sigma} \bar{\phi}}
	{1+ \vert \phi \vert^{2}}
\right)	
S^{\rho \sigma}	
\, .
\label{Finsler-metric-spinning}
\end{align}

One can generalize the duality among the Finsler geometries from the spin-$0$ particle to the spin-$\frac{1}{2}$ particle as follows
\begin{align}
& g^{(\frac{1}{2})}_{\mu \nu}  
[ x, \dot{x}, \psi, \dot{\psi}; e, \chi , A]
\to 
\tilde{g}^{(\frac{1}{2})}_{\mu \nu} 
[y, \dot{y}, \xi, \dot{\xi}; f, \kappa , C ] \, : 
\nonumber
\\
&
\{ x \to y \, , 
e \to f \, , 
\psi \to \zeta \, ,
\chi \to \kappa \, ,
\phi \to - \theta \, , 
m \to M \, , 
q \to - k 
\}
\, .    
\label{g-tilde-g-map-spinning}
\end{align}
These maps leave the Lagrangian invariant but change the description of the spinning particle in the electromagnetic field $A_{\mu}$ to a formally magnetic charged spinning particle in the dual field ${C}_{\mu}$. 

Similarly, the calculation of the Cartan tensor produces the following result
\begin{align}
C^{(\frac{1}{2})}_{\mu \nu \rho} & = 
+ \frac{q}{e}
\left[
\frac{
\eta_{\mu \nu} 
\left(
\bar{\phi} \partial_{\rho} \phi - \phi \partial_{\rho} \bar{\phi}
\right)
+ 2\eta_{\rho (\mu} 
\left(
\bar{\phi} \partial_{\nu)} \phi - \phi \partial_{\nu)} \bar{\phi}
\right) 
}{1+ \vert \phi \vert^{2}}
\right] 
\nonumber
\\
& 
+ \frac{1}{e^2}
\left[
\eta_{\mu \nu} \eta_{\rho  \sigma } + 
\eta_{\mu \rho} \eta_{\nu \sigma} +
\eta_{\mu \sigma} \eta_{\nu \rho}
\right] \dot{x}^{\sigma}
-\frac{i}{2e} \chi \eta_{\mu \nu }
\psi_{\rho} 
\, .
\label{Cartan-Finsler-metric-spinning}
\end{align} 
We can see that the above relations reduce to their spin-$0$ particle counterparts in the absence of the spinning variables and can be simplified by choosing convenient reparametrizations of the world-line. For example, the metric $g^{(\frac{1}{2})}_{\mu \nu}$ from the equation (\refeq{Finsler-metric-spinning}) simplifies considerably in the proper-time gauge, that is for $e = 1/m$ and $\chi  = 0$. Other gauges are possible, and they are obtained by using the equations of motion.

The angular metric tensor and the Finsler geodesic equations of motion in the Ra\~{n}ada background can be computed from their respective definitions. The corresponding relations show that the dependency of these objects on the topological data of the background field, encoded by the function $\phi$, is highly nonlinear. Since the formulas are quite large, they are presented in the Appendix C.

\section{Conclusions}

In summary, we have constructed the Finsler geometries and calculated the fundamental objects for the kinematic spaces of the spin-$0$ and spin-$\frac{1}{2}$ particles in an arbitrary electromagnetic field in the most general formulation of the particle action in which all symmetries are manifest. We have given a simple method to calculate the metric tensor, the angular metric tensor, the Cartan tensor and the geodesic equations 
that are all parametrized by the world-sheet einbein and its supersymmetric partner that are associated to the gauge reparametrization transformations and to the electromagnetic potentials. Due to this parametrization, the Finsler structure is actually a set of Finsler spaces. Our calculations show that, in general, the inverse metric and the objects derived from it are non-polynomial and non-linear in all variables. Next, we have applied this method to calculate the fundamental tensors of the Finsler geometry for the particle in motion in the topological Ra\~{n}ada field. Here, we have showed that there is a duality map among Finsler geometries induced by the electromagnetic duality. We have given a simple interpretation of this map as a transformation between the original particle in the electromagnetic field and a formal magnetically charged particle moving in the dual field and we have exemplified its application by calculating the geodesic equation in the dual Finsler geometry in the spin-$0$ case.

Working with the full symmetries of the particle action is important for determining the full set of Finsler spaces as the gauge fixed actions, e. g. the proper-time action, misses spaces from this set. However, for practical applications such as solving the geodesic equation and quantization, it is necessary to fix the reparametrization and the electromagnetic gauges by which a particular Finsler geometry is also selected from the set of allowed geometries. Nevertheless, even if the gauge symmetries are fixed, it is difficult to solve the geodesic equation due to the non-polynomial and non-linear terms present in it.

The results obtained here open up a rich set of research lines for future investigations. One interesting problem is whether the dual map between the Finsler geometries has generators and, more generally, the structure of the dualities among the geometries and their relation with physics. Another problem is to study the Finsler geometries corresponding to the polynomial action which implies including the constraints in the fundamental Finsler function. New generalizations of the Ra\~{n}ada fields have been presented recently in the literature, see e. g. \cite{Thompson:2014pta,Kopinski:2017nvp,Vancea:2017tmx,Silva:2018ule} for topological solutions in the presence of the gravitational field and 
\cite{deKlerk:2017qvq,Goulart:2016orx,Hoyos:2015bxa,Alves:2017zjt,Alves:2017ggb,Nastase:2018gjs} for generalization to the non-linear electrodynamics. It should be interesting to generalize the construction presented in this paper to determine the Finsler geometries associated to these systems. In particular, it would be interesting to see whether any of these solutions can be connected to the $b$-Finsler geometry \cite{Kostelecky:2011,Foster:2015,Schreck:2015}. While the results presented here have been obtained by applying classical analytic methods of differential geometry, it is certainly interesting to implement a computer assisted algorithm for investigating the aforementioned problems and other. For example, the geodesic equations obtained in this paper are highly non-linear. Therefore, from the point of view of the applications of these results, it is important to investigate the solubility of the geodesic equations and their solutions either analytically as well as numerically.

\section*{Acknowledgements} I. V. V. would like to acknowledge N. Berkovits for hospitality at ICTP-SAIFR where part of this work was done and to H. Nastase for discussions. Also, we would like to thank to an anonymous referee whose consistent comments have helped us to improve greatly the quality of the text.

\section*{Appendix A: Basic concepts of Finsler geometry}
\setcounter{equation}{0}
\renewcommand{\theequation}{A.\arabic{equation}} 

In this Appendix, we review some basic concepts and relations of Finsler geometry that have been used above. For more details we refer to \cite{Rund:1959}.

Let $\mathcal{M}$ be a real manifold of dimension $n$ endowed with a real scalar function $F[x,y]$ where $x \in \mathcal{M}$ and $y \in \mathcal{T}_{x}(\mathcal{M})$ which is the tangent space at $x$. The function $F[x,y]$, called the \emph{fundamental function} of the Finsler space or the \emph{Finsler metric function}, is required to have the following local properties: 
\begin{itemize}

\item[i)] $F[x,y]$ is smooth with respect to $x$ and $y$. 

\item[ii)] There is a conical submanifold $\mathcal{M}^{\vee}_{x}(\mathcal{M}) \subset \mathcal{T}_{x}(\mathcal{M})$ such that:  

$F[x,y]>0$ and $\det \,[\partial^{2}_{yy} \, F^2 [x,y]] \neq 0$ 
for any $y \in \mathcal{V}_{x}(\mathcal{M})$.

\item[iii)] $F[x,y]$ is homogeneous of degree one in $y$, i. e. $F(x,\lambda y) = \lambda F[x,y]$ for any $\lambda > 0$ and any $y \in \mathcal{M}^{\vee}_{x}(\mathcal{M})$.

\end{itemize}
An \emph{n-dimensional Finsler space} is a triple 
$( \mathcal{M}, \mathcal{M}^{\vee}_{x}(\mathcal{M}), F[x,y] )$ with the properties i) - iii) above. From these properties, one can readily write the fundamental objects of the Finsler geometry in local coordinates $\{ x^{i} \}$, where $i = \overline{1,n}$ on $\mathcal{M}$. The first such objects are the \emph{Finsler metric tensor}  
$g_{ij}[x,y]$, the \emph{unit tangent vector} $l_i$ and the \emph{angular metric tensor} $h_{ij}[x,y]$ defined by the following relations
\begin{equation}
g_{ij}[x,y] := \frac{1}{2} \frac{\partial^2 F^2 [x,y]}{\partial y^i \partial y^j}
\, ,
\quad
l_{i}[x,y] := \frac{y_i}{F[x,y]}
\, ,
\quad
h_{ij}[x,y] := g_{ij}[x,y] - l_{i}[x,y]l_{j}[x,y]
\, .
\label{metric-unit-angular-Finsler-def}
\end{equation}
The \emph{Cartan tensor} can also be expressed in terms of the fundamental function as follows
\begin{equation}
C_{ijk}[x,y] : = \frac{1}{2} \frac{\partial g_{ij}[x,y]}{\partial y^k}
= \frac{1}{4} \frac{\partial^3 F^{2}[x,y]}{\partial y^i \partial y^j \partial y^k}
\, .
\label{Cartan-tensor}
\end{equation}
Of interest in the Finsler geometry of the kinematic space of a physical particle is the \emph{Finsler geodesic equation} as the variational problem of the equation of motion can be reduced to the geodesic problem \cite{Rund:1959}. By generalizing the definition from the Riemann geometry, the Finsler geodesic is defined as the stationary solution to the variational problem
\begin{equation}
\delta I [\gamma ] = \delta \int_{\gamma} F(x, y = dx ) = 0
\, ,
\label{geodesic-variational-Finsler-definition}
\end{equation}
where $\gamma : [\tau_1 , \tau_2 ] \subset \mathbb{R} \to \mathcal{M}$ is a smooth curve parametrized by $\tau$ with fixed endpoints $x_{1} = \gamma (\tau_1 )$ and $x_2 = \gamma(\tau_2)$. The Finsler geodesic equation takes a similar form to the geodesic equation from the Riemann geometry when expressed in terms of the \emph{Finsler-Christoffel symbols} defined by the usual relations
\begin{equation}
{\Gamma^{i}}_{j k} := \frac{1}{2} g^{i r}
\left(
 \frac{\partial g_{r k}}{\partial x^{j}} +
 \frac{\partial g_{r j}}{\partial x^{k}}-
 \frac{\partial g_{j k}}{\partial x^{r}}
\right)
\, .
\label{Finsler-Christoffel-symbols-definition}
\end{equation}
Then, the variational problem (\refeq{geodesic-variational-Finsler-definition}) generates the Finsler geodesic equation
\begin{equation}
\frac{d^2 x^{i}}{d\tau^{2}} + {\Gamma^{i}}_{j k} 
\frac{d x^{j}}{d \tau} 
\frac{d x^{k}}{d \tau}
- \frac{d x^{i}}{d \tau} \frac{d}{d \tau} 
	\left[ 
	\ln(F)	
	\right]
= 0
\, .
\label{geodesic-equation-Finsler-definition}
\end{equation}
As is well known, the similarity between the Riemann geometry and the Finsler geometry is not accidental. The Riemann geometry is a particular case of the Finsler geometry for the metric tensor $g[x,y]$ independent on $y$ which implies that $C_{ijk} = 0$.

\section*{Appendix B: Dual geodesic equations}
\setcounter{equation}{0}
\renewcommand{\theequation}{B.\arabic{equation}}
 
Here, we present the angular metric tensor and the Finsler geodesic equation for the spinless particle in the Ra\~{n}ada background.
The angular metric tensor has the following form
\begin{align}
h_{\mu \nu} [x, \dot{x}, e; A ] & = 
\frac{m^2}{2 } \eta_{\mu \nu} -   
\frac{q^2 a}{ 16 \pi^2}
\left(
\frac{\bar{\phi} \partial_{\mu} \phi - \phi \partial_{\mu} \bar{\phi}}{1+ \vert \phi \vert^{2}}
\right)
\left(
\frac{\bar{\phi} \partial_{\nu} \phi - \phi \partial_{\nu} \bar{\phi}}{1+ \vert \phi \vert^{2}}
\right)
\nonumber
\\
& + \frac{q}{e}
\left[
\frac{
\eta_{\mu \nu} 
\left(
\bar{\phi} \partial_{\rho} \phi - \phi \partial_{\rho} \bar{\phi}
\right)
+ 2\eta_{\rho (\mu} 
\left(
\bar{\phi} \partial_{\nu)} \phi - \phi \partial_{\nu)} \bar{\phi}
\right) 
}{1+ \vert \phi \vert^{2}}
\right] \dot{x}^{\rho}
\, ,
\nonumber
\\
& - 
\left\{
\left[
	\frac{1}{4e^2} 
	\left( \dot{x}^{2} \right)^2
	+ \frac{1}{4} e^2 m^4 
	- 
		\frac{a q^2}{16 \pi^2 } 
		\left(
		\frac{\bar{\phi} \partial_{\rho} \phi - \phi \partial_{\rho}
		\bar{\phi}}{1+ \vert \phi \vert^{2}}
		\right) 
		\left(
		\frac{\bar{\phi} \partial_{\sigma} \phi - \phi \partial_{\sigma}
		\bar{\phi}}{1+ \vert \phi \vert^{2}}
		\right) \dot{x}^{\rho} \dot{x}^{\sigma}
	\right]^{-2}
\right.
\nonumber
\\
& +
\left.
\frac{m^2}{2} \dot{x}^2
+
	\frac{q\sqrt{a}}{8 \pi e i} 
	\left(
		\frac{\bar{\phi} \partial_{\rho} \phi - \phi \partial_{\rho}
		\bar{\phi}}{1+ \vert \phi \vert^{2}}
	\right) \dot{x}^{\rho} \dot{x}^{2}
+
	\frac{q e\sqrt{a}}{8 m^2 \pi e i} 
	\left(
		\frac{\bar{\phi} \partial_{\rho} \phi - \phi \partial_{\rho}
		\bar{\phi}}{1+ \vert \phi \vert^{2}}
	\right) \dot{x}^{\rho} 
\right\}
\dot{x}_{\mu} \dot{x}_{\nu}
\, .
\label{angular-momentum-tensor-Ranada}
\end{align}
One can calculate the Finsler geodesic equations in the spaces 
$\mathfrak{X}[x,\dot{x}; e , A]$ and 
$\mathfrak{Y}[y,\dot{y}; f , C]$. The first equation is obtained from the 
equations (\refeq{A-potentials}) and (\refeq{geodesic-equation-Finsler-final}) and has the following form
\begin{align}
\frac{d^2 {x}^{\mu}}{d \tau^2 }  & +
\frac{e}{2}
\left\{
\left[
\frac{1}{2e} \dot{x}^{2}
+ \frac{1}{2} e m^2 + 
\frac{q \sqrt{a}}{4 \pi i} 
	\left(
	\frac{\bar{\phi} \partial_{\rho} \phi - \phi \partial_{\rho} 					\bar{\phi}}{1+ \vert \phi \vert^{2}}
	\right)
	\dot{x}^{\rho}
\right]
\right\}^{-1}
\nonumber
\\ 
& \times 
\left\{ 
\eta^{\mu \nu} - 
\left[
\frac{1}{e^2} \dot{x}^{\mu} \dot{x}^{\nu} 
+
\frac{q \sqrt{a}}{4 \pi i}
\left[
\left(
\frac{\bar{\phi} \partial^{\mu} \phi - \phi \partial^{\mu} \bar{\phi}}{1+ \vert \phi \vert^{2}}
\right) \dot{x}^{\nu}
+
\left(
\frac{\bar{\phi} \partial^{\nu} \phi - \phi \partial^{\nu} \bar{\phi}}{1+ \vert \phi \vert^{2}}
\right) \dot{x}^{\mu}
\right]
\right]
\right.
\nonumber
\\
& \times
\left.
	\left[
	\frac{1}{e^2} \dot{x}^2  +
	\frac{q \sqrt{a}}{2 \pi i}
		\left(
		\frac{\bar{\phi} \partial_{\rho} \phi - \phi \partial_{\rho} 					\bar{\phi}}{1+ \vert \phi \vert^{2}}
		\right) \dot{x}^{\rho}
	- 
	\frac{q^2 a}{16 \pi^2}
	\left(
	\frac{\bar{\phi} \partial_{\rho} \phi - \phi \partial_{\rho} 					\bar{\phi}}{1+ \vert \phi \vert^{2}}
	\right)
		\left(
		\frac{\bar{\phi} \partial^{\rho} \phi - \phi \partial^{\rho} 					\bar{\phi}}{1+ \vert \phi \vert^{2}}
		\right)	
	\right.	
\right.
\nonumber
\\
& +
\left.
	\left.	
\frac{1}{2e^2} \dot{x}^{2}
+ \frac{1}{2} m^2 + 
\frac{q \sqrt{a}}{4 \pi i e} 
		\left(
		\frac{\bar{\phi} \partial_{\rho} \phi - \phi \partial_{\rho} 					\bar{\phi}}{1+ \vert \phi \vert^{2}}
		\right)
	\dot{x}^{\rho}	
	\right]^{-1}
\right\}
\nonumber
\\
& \times
\left\{
\frac{1}{e}
	\left\{ 
	2 x_{\nu} \boldsymbol{\partial}
		\left[
		\frac{1}{2e} \dot{x}^{2}
		+ \frac{1}{2} e m^2 + 
		\frac{q \sqrt{a}}{4 \pi i} 
			\left(
			\frac{\bar{\phi} \partial_{\rho} \phi - \phi \partial_{\rho} 					\bar{\phi}}{1+ \vert \phi \vert^{2}}
			\right)
			\dot{x}^{\rho}
		\right]
	\right.	
\right.	
\nonumber
\\
& -
\left.
	\left.
	\dot{x}^{2} \partial_{\nu} 
	\left[
		\frac{1}{2e} \dot{x}^{2}
		+ \frac{1}{2} e m^2 + 
		\frac{q \sqrt{a}}{4 \pi i} 
			\left(
			\frac{\bar{\phi} \partial_{\rho} \phi - \phi \partial_{\rho} 					\bar{\phi}}{1+ \vert \phi \vert^{2}}
			\right)
			\dot{x}^{\rho}
		\right]
	\right\}
\right.
\nonumber
\\
&+
\left. 
    \frac{\sqrt{a}}{4 \pi i}	
	\dot{x}^{\rho}\partial_{\rho} 
	\left[	
		\left( 
		\frac{1}{e} \dot{x}^2 +
		\dot{x}^{\sigma}
   		\frac{\bar{\phi} \partial_{\sigma} \phi 
    	- \phi \partial_{\sigma} \bar{\phi}}{1+ \vert \phi \vert^{2}}
    	\right)  	
    	\left(
    	\frac{1}{e} \dot{x}_{\nu}
		+ \frac{q \sqrt{a}}{4 \pi i}
			\left(
			\frac{\bar{\phi} \partial_{\nu} \phi - \phi \partial_{\nu} 					\bar{\phi}}{1+ \vert \phi \vert^{2}}
			\right)
		\right)
	\right]
\right.
\nonumber
\\
& -
\left.	
	 \partial_{\nu} 
	 \left[ 
	 \frac{1}{e^2} \dot{x}^2  
		+
		\frac{q \sqrt{a}}{2 \pi i}
		\left(
		\frac{\bar{\phi} \partial_{\lambda} \phi - \phi
	 	\partial_{\lambda} \bar{\phi}}{1+ \vert \phi \vert^{2}}
		\right) \dot{x}^{\lambda}
	- 
	\frac{q^2 a}{16 \pi^2}
	\left(
	\frac{\bar{\phi} \partial_{\lambda} \phi - \phi \partial_{\lambda} 			\bar{\phi}}{1+ \vert \phi \vert^{2}}
	\right)
	\left(
	\frac{\bar{\phi} \partial^{\lambda} \phi - \phi \partial^{\lambda} 					\bar{\phi}}{1+ \vert \phi \vert^{2}}
	\right)
	\right]
\right\}
\nonumber
\\
&
= \dot{x}^{\mu} \frac{d}{d \tau} 
\left\{ 
\ln 
	\left[
	\frac{1}{2e} \dot{x}^{2} + \frac{1}{2} e m^2 + 
	\frac{q \sqrt{a}}{4 \pi i} 
		\left(
		\frac{\bar{\phi} \partial_{\rho} \phi - \phi \partial_{\rho} 				\bar{\phi}}{1+ \vert \phi \vert^{2}}
		\right)
		\dot{x}^{\rho}
	\right]	
\right\}
\, .
\label{geodesic-equation-Finsler-Ranada}
\end{align} 
The geodesic equation is non-linear and non-polynomial in all its variables. The geodesic equation can be calculated in two ways, either from its definition or by applying the duality map \refeq{g-tilde-g-map}) to the geodesic equation (\refeq{geodesic-equation-Finsler-Ranada}).
The result is the following equation
\begin{align}
\frac{d^2 {y}^{\mu}}{d \tau^2 }  & +
\frac{f}{2}
\left\{
\left[
\frac{1}{2f} \dot{y}^{2}
- \frac{1}{2} f M^2 + 
\frac{k \sqrt{a}}{4 \pi i} 
	\left(
	\frac{\bar{\theta} \partial_{\rho} \theta - \theta \partial_{\rho} 					\bar{\theta}}{1+ \vert \theta \vert^{2}}
	\right)
	\dot{y}^{\rho}
\right]
\right\}^{-1}
\nonumber
\\ 
& \times 
\left\{ 
\eta^{\mu \nu} - 
\left[
\frac{1}{f^2} \dot{y}^{\mu} \dot{y}^{\nu} 
-
\frac{k \sqrt{a}}{4 \pi i}
\left[
\left(
\frac{\bar{\theta} \partial^{\mu} \theta - \theta \partial^{\mu} \bar{\theta}}{1+ \vert \theta \vert^{2}}
\right) \dot{y}^{\nu}
+
\left(
\frac{\bar{\theta} \partial^{\nu} \theta - \theta \partial^{\nu} \bar{\theta}}{1+ \vert \theta \vert^{2}}
\right) \dot{y}^{\mu}
\right]
\right]
\right.
\nonumber
\\
& \times
\left.
	\left[
	\frac{1}{f^2} \dot{y}^2  -
	\frac{k \sqrt{a}}{2 \pi i}
		\left(
		\frac{\bar{\theta} \partial_{\theta} \theta - \theta						\partial_{\rho}	\bar{\theta}}{1+ \vert \theta \vert^{2}}
		\right) \dot{y}^{\rho}
	- 
	\frac{k^2 a}{16 \pi^2}
	\left(
	\frac{\bar{\theta} \partial_{\rho} \theta - \theta \partial_{\rho} 					\bar{\theta}}{1+ \vert \theta \vert^{2}}
	\right)
		\left(
		\frac{\bar{\theta} \partial^{\rho} \theta - \theta
		\partial^{\rho} \bar{\theta}}{1+ \vert \theta \vert^{2}}
		\right)	
	\right.	
\right.
\nonumber
\\
& +
\left.
	\left.	
\frac{1}{2f^2} \dot{y}^{2}
+ \frac{1}{2} M^2 -
\frac{k \sqrt{a}}{4 \pi i f} 
		\left(
		\frac{\bar{\theta} \partial_{\rho} \theta - \theta
		\partial_{\rho} \bar{\theta}}{1+ \vert \theta \vert^{2}}
		\right)
	\dot{y}^{\rho}	
	\right]^{-1}
\right\}
\nonumber
\\
& \times
\left\{
\frac{1}{f}
	\left\{ 
	2 y_{\nu} \boldsymbol{\partial}
		\left[
		\frac{1}{2f} \dot{y}^{2}
		+ \frac{1}{2} f M^2 -
		\frac{k \sqrt{a}}{4 \pi i} 
			\left(
			\frac{\bar{\theta} \partial_{\rho} \theta - \theta							\partial_{\rho}	\bar{\theta}}{1+ \vert \theta \vert^{2}}
			\right)
			\dot{y}^{\rho}
		\right]
	\right.	
\right.	
\nonumber
\\
& -
\left.
	\left.
	\dot{y}^{2} \partial_{\nu} 
	\left[
		\frac{1}{2f} \dot{y}^{2}
		+ \frac{1}{2} f M^2 - 
		\frac{k \sqrt{a}}{4 \pi i} 
			\left(
			\frac{\bar{\theta} \partial_{\rho} \theta - \theta 							\partial_{\rho} \bar{\theta}}{1+ \vert \theta \vert^{2}}
			\right)
			\dot{y}^{\rho}
		\right]
	\right\}
\right.
\nonumber
\\
&+
\left. 
    \frac{\sqrt{a}}{4 \pi i}	
	\dot{y}^{\rho}\partial_{\rho} 
	\left[	
		\left( 
		\frac{1}{f} \dot{y}^2 +
		\dot{y}^{\sigma}
   		\frac{\bar{\theta} \partial_{\sigma} \theta 
    	- \theta \partial_{\sigma} \bar{\theta}}{1+ \vert \theta					\vert^{2}}
    	\right)  	
    	\left(
    	\frac{1}{f} \dot{y}_{\nu}
		- \frac{k \sqrt{a}}{4 \pi i}
			\left(
			\frac{\bar{\theta} \partial_{\nu} \theta - \theta 							\partial_{\nu} \bar{\theta}}{1+ \vert \theta \vert^{2}}
			\right)
		\right)
	\right]
\right.
\nonumber
\\
& -
\left.	
	 \partial_{\nu} 
	 \left[ 
	 \frac{1}{f^2} \dot{y}^2  
		-
		\frac{k \sqrt{a}}{2 \pi i}
		\left(
		\frac{\bar{\theta} \partial_{\lambda} \theta - 
		\theta \partial_{\lambda} \bar{\phi}}{1+ \vert \theta \vert^{2}}
		\right) \dot{y}^{\lambda}
	- 
	\frac{k^2 a}{16 \pi^2}
	\left(
	\frac{\bar{\theta} \partial_{\lambda} \theta - \theta 
	\partial_{\lambda} 	\bar{\phi}}{1+ \vert \theta \vert^{2}}
	\right)
	\left(
	\frac{\bar{\theta} \partial^{\lambda} \theta - 
	\theta \partial^{\lambda} \bar{\phi}}{1+ \vert \theta \vert^{2}}
	\right)
	\right]
\right\}
\nonumber
\\
&
= \dot{y}^{\mu} \frac{d}{d \tau} 
\left\{ 
\ln 
	\left[
	\frac{1}{2f} \dot{y}^{2} + \frac{1}{2} f M^2 - 
	\frac{k \sqrt{a}}{4 \pi i} 
		\left(
		\frac{\bar{\theta} \partial_{\rho} \theta - \theta
		\partial_{\rho}	\bar{\theta}}{1+ \vert \theta \vert^{2}}
		\right)
		\dot{y}^{\rho}
	\right]	
\right\}
\, .
\label{geodesic-equation-Finsler-Ranada-dual}
\end{align} 
Since the duality map only requires a simple substitution of variables, it is provides a  faster way to construct the geometrical objects from the dual geometry. The geodesic equations from above illustrate the advantage of knowing the duality map from the computational point of view. Since the system formed by the electrically and magnetically charged particles support the two dual geometries, the physical content is unchanged by performing the dual transformations between these geometries, which is the content of the equivalence of the geometries under the duality map.

\section*{Appendix C: Angular metric and geodesic equation for spinning particle}
\setcounter{equation}{0}
\renewcommand{\theequation}{C.\arabic{equation}} 

Here, we present the angular metric tensor and the Finsler geodesic equation on the space $\mathfrak{X}_{(\frac{1}{2})}[ x, \dot{x}, \psi, \dot{\psi}; e, \chi , A]$. 

The field angular metric tensor for the spinning particle in the field $A_{\mu}$ has been calculated from its definition. After some simplifications, it can be put under the following compact form 
\begin{align}
h^{(\frac{1}{2})}_{\mu \nu}  & = 
\frac{m^2}{2 } \eta_{\mu \nu} -   
\frac{q^2 a}{ 16 \pi^2}
\left(
\frac{\bar{\phi} \partial_{\mu} \phi - \phi \partial_{\mu} \bar{\phi}}{1+ \vert \phi \vert^{2}}
\right)
\left(
\frac{\bar{\phi} \partial_{\nu} \phi - \phi \partial_{\nu} \bar{\phi}}{1+ \vert \phi \vert^{2}}
\right)
\nonumber
\\
& + \frac{q}{e}
\left[
\frac{
\eta_{\mu \nu} 
\left(
\bar{\phi} \partial_{\rho} \phi - \phi \partial_{\rho} \bar{\phi}
\right)
+ 2\eta_{\rho (\mu} 
\left(
\bar{\phi} \partial_{\nu)} \phi - \phi \partial_{\nu)} \bar{\phi}
\right) 
}{1+ \vert \phi \vert^{2}}
\right] \dot{x}^{\rho}
\nonumber
\\
& 
+ \frac{1}{2 e^2}
\left[
\eta_{\mu \nu} \dot{x}^2 + 2 \eta_{\mu \rho} \eta_{\nu \sigma}
\dot{x}^{\rho} \dot{x}^{\sigma}
\right]
+ \frac{i}{2 e} \eta_{\mu \nu}
\left(
\dot{\psi}_{\rho} \psi^{\rho} - \dot{\psi}_5 \psi_5
\right)
\nonumber
\\
&
-\frac{i}{2e} \chi
\left[
\frac{\sqrt{a}\, e}{2\pi i}
\left(
\frac{\psi_{(\mu} \bar{\phi} \partial_{\nu)} \phi - 
	\psi_{(\mu}\phi \partial_{\nu)} \bar{\phi}}
	{1+ \vert \phi \vert^{2}}
\right) 
+ 
\eta_{\mu \nu }
\left(
\psi_{\rho} \dot{x}^{\rho}
- 2  m \psi_5  
\right)
\right]
\nonumber
\\
& +
\eta_{\mu \nu}
\frac{q \sqrt{a}}{8\pi i}
\left(
\frac{\partial_{\rho} \bar{\phi} \partial_{\sigma} \phi - 
	\partial_{\rho}\phi \partial_{\sigma} \bar{\phi}}
	{1+ \vert \phi \vert^{2}}
\right)	
S^{\rho \sigma}	
\nonumber
\\
& - 
\left[
\frac{1}{2e} \dot{x}^{2} + \frac{1}{2} e m^2
+ \frac{i}{2} 
\left(
\dot{\psi}_{\lambda} \psi^{\lambda} - \dot{\psi}_{5} \psi_{5}
\right)
- \frac{i}{2e} \chi  \psi_{\lambda} \dot{x}^{\lambda}
+ im \chi \psi_{5}
\right.
\nonumber
\\
& +
\left.
\frac{q \sqrt{a}}{4 \pi i} 
	\left(
	\frac{\bar{\phi} \partial_{\rho} \phi - \phi \partial_{\rho} 					\bar{\phi}}{1+ \vert \phi \vert^{2}}
	\right)
	\dot{x}^{\rho}
+ 
\frac{i q \sqrt{a}}{4 \pi i} e
	\left(
	\frac{\partial_{\rho} \bar{\phi} \partial_{\sigma} \phi - 
	\partial_{\rho}\phi \partial_{\sigma} \bar{\phi}}
	{1+ \vert \phi \vert^{2}}
	\right)
	S^{\rho \sigma} 
\right]^{-2}
\dot{x}_{\mu} \dot{x}_{\nu}
\, .
\label{angular-momentum-tensor-spinning-particle}
\end{align}
The Finsler geodesic equation of the spinning particle, calculated from its definition, has the following form
\begin{align}
\frac{d^2 {x}^{\mu}}{d \tau^2 }  & +
\frac{e}{2}
\left\{
\left[
\frac{1}{2e} \dot{x}^{2}
+ \frac{1}{2} e m^2 + 
\frac{q \sqrt{a}}{4 \pi i} 
	\left(
	\frac{\bar{\phi} \partial_{\lambda} \phi - \phi \partial_{\lambda} 					\bar{\phi}}{1+ \vert \phi \vert^{2}}
	\right)
	\dot{x}^{\lambda}
	+ \frac{i}{2}
	\left(
	\dot{\psi}_{\lambda} \psi^{\lambda} - \dot{\psi}_{5} \psi^{5}
	\right) 
	- \frac{i}{2} \chi \theta_{\lambda} \dot{x}^{\lambda}
	+i n \chi \theta_5 
\right.
\right.
\nonumber
\\
& +  
\left.
\left.
\frac{q^2 \sqrt{a}}{4\pi i} e
\left(
\frac{\partial_{\rho} \bar{\phi} \partial_{\sigma} \phi - 
	\partial_{\rho}\phi \partial_{\sigma} \bar{\phi}}
	{1+ \vert \phi \vert^{2}}
\right)
S^{\rho \sigma}		
\right]
\right\}^{-1}
\nonumber
\\ 
& \times 
\left\{ 
\eta^{\mu \nu} - 
	\left[
	\frac{1}{e^2} \dot{x}^{2} 
	+
	\frac{q \sqrt{a}}{4 \pi i}
		\left[
		\left(
			\frac{\bar{\phi} \partial^{\mu} \phi - \phi \partial^{\mu}
			\bar{\phi}}{1+ \vert \phi \vert^{2}}
		\right) \dot{x}^{\nu}
		+
		\left(
		\frac{\bar{\phi} \partial^{\nu} \phi - \phi \partial^{\nu} 					\bar{\phi}}{1+ \vert \phi \vert^{2}}
		\right) \dot{x}^{\mu}
		\right.
	\right.
\right.
\nonumber
\\
& - 
\left.
	\left.
		\left.
		i \chi \psi^{(\mu} 
		\left(
		\frac{1}{e} \dot{x}^{\nu )} + 
		\frac{q^2 \sqrt{a}}{4 \pi i} 
		\left(
		\frac{\bar{\phi} \partial^{\nu)} \phi - \phi \partial^{\nu)} 					\bar{\phi}}{1+ \vert \phi \vert^{2}}
		\right)		
		\right)
		\right]
	\right]
\right.
\nonumber
\\
& \times
\left.
	\left[
	\frac{1}{e^2} \dot{x}^2  +
	\frac{q \sqrt{a}}{2 \pi i}
		\left(
		\frac{\bar{\phi} \partial_{\rho} \phi - \phi \partial_{\rho} 					\bar{\phi}}{1+ \vert \phi \vert^{2}}
		\right) \dot{x}^{\rho}
	- 
	\frac{q^2 a}{16 \pi^2}
	\left(
	\frac{\bar{\phi} \partial_{\rho} \phi - \phi \partial_{\rho} 					\bar{\phi}}{1+ \vert \phi \vert^{2}}
	\right)
		\left(
		\frac{\bar{\phi} \partial^{\rho} \phi - \phi \partial^{\rho} 					\bar{\phi}}{1+ \vert \phi \vert^{2}}
		\right)	
	\right.	
\right.
\nonumber
\\
& -
\left.
	\left.
	\frac{i}{e} \chi \eta^{\rho \sigma}\psi_{(\rho}
		\left(
		\dot{x}_{\sigma)} + 
		\frac{q^2 \sqrt{a}}{4 \pi i}
		\left( 
		\frac{\bar{\phi} \partial_{\sigma)} \phi - \phi \partial_{\sigma 			)}\bar{\phi}}{1+ \vert \phi \vert^{2}}
		\right)
		\right)
	\right.	
\right.
\nonumber
\\
& +
\left.
	\left.	
\frac{1}{2e^2} \dot{x}^{2}
+ \frac{1}{2} m^2 + 
\frac{q \sqrt{a}}{4 \pi i e} 
		\left(
		\frac{\bar{\phi} \partial_{\rho} \phi - \phi \partial_{\rho} 					\bar{\phi}}{1+ \vert \phi \vert^{2}}
		\right)
		\dot{x}^{\rho}
		 +
		\frac{i}{2}
		\left(
		\dot{\psi}_{\lambda} \psi^{\lambda} - \dot{\psi}_{5} \psi^{5}
		\right) 
		- \frac{i}{2} \chi \theta_{\lambda} \dot{x}^{\lambda} 
		+i n \chi \theta_5 	
	\right.
\right.
\nonumber
\\
& + 
\left.
	\left.
\frac{q^2 \sqrt{a}}{4\pi i} e
		\left(
		\frac{\partial_{\rho} \bar{\phi} \partial_{\sigma} \phi - 
		\partial_{\rho}\phi \partial_{\sigma} \bar{\phi}}
		{1+ \vert \phi \vert^{2}}
		\right)
		S^{\rho \sigma}		
	\right]^{-1}
\right\}
\nonumber
\\ 
& \times
\left\{
\frac{1}{e}
	\left\{ 
	2 x_{\nu} \boldsymbol{\partial}
		\left[
		\frac{1}{2e} \dot{x}^{2}
		+ \frac{1}{2} e m^2 + 
		\frac{q \sqrt{a}}{4 \pi i} 
			\left(
			\frac{\bar{\phi} \partial_{\rho} \phi - \phi \partial_{\rho} 					\bar{\phi}}{1+ \vert \phi \vert^{2}}
			\right)
			\dot{x}^{\rho}
		\right]
	\right.	
\right.
\nonumber
\\
& -
\left.
	\left.
	\dot{x}^{2} \partial_{\nu} 
	\left[
		\frac{1}{2e} \dot{x}^{2}
		+ \frac{1}{2} e m^2 + 
		\frac{q \sqrt{a}}{4 \pi i} 
			\left(
			\frac{\bar{\phi} \partial_{\rho} \phi - \phi \partial_{\rho} 					\bar{\phi}}{1+ \vert \phi \vert^{2}}
			\right)
			\dot{x}^{\rho}
		\right]
	\right\}
\right.
\nonumber
\\
&+
\left. 
    \frac{\sqrt{a}}{4 \pi i}	
	\dot{x}^{\rho}\partial_{\rho} 
	\left[	
		\left( 
		\frac{1}{e} \dot{x}^2 +
		\dot{x}^{\sigma}
   		\frac{\bar{\phi} \partial_{\sigma} \phi 
    	- \phi \partial_{\sigma} \bar{\phi}}{1+ \vert \phi \vert^{2}}
    	\right)  	
    	\left(
    	\frac{1}{e} \dot{x}_{\nu}
		+ \frac{q \sqrt{a}}{4 \pi i}
			\left(
			\frac{\bar{\phi} \partial_{\nu} \phi - \phi \partial_{\nu} 					\bar{\phi}}{1+ \vert \phi \vert^{2}}
			\right)
		\right)
	\right]
\right.
\nonumber
\\
& -
\left.	
	 \partial_{\nu} 
	 \left[ 
	 \frac{1}{e^2} \dot{x}^2  
		+
		\frac{q \sqrt{a}}{2 \pi i}
		\left(
		\frac{\bar{\phi} \partial_{\lambda} \phi - \phi
		\partial_{\lambda} \bar{\phi}}{1+ \vert \phi \vert^{2}}
		\right) \dot{x}^{\lambda}
	- 
	\frac{q^2 a}{16 \pi^2}
	\left(
	\frac{\bar{\phi} \partial_{\lambda} \phi - \phi \partial_{\lambda} 					\bar{\phi}}{1+ \vert \phi \vert^{2}}
	\right)
	\left(
	\frac{\bar{\phi} \partial^{\lambda} \phi - \phi \partial^{\lambda} 					\bar{\phi}}{1+ \vert \phi \vert^{2}}
	\right)
    \right.
\right.
\nonumber
\\
& +
\left.
	\left.
	\frac{q^2 \sqrt{a}}{4\pi }
		\left(
		2 \partial_{\rho}			
			\left( 
			\frac{\partial_{\omega} \bar{\phi} \partial_{\sigma} \phi - 
			\partial_{\omega}\phi \partial_{\sigma} \bar{\phi}}
			{1+ \vert \phi \vert^{2}}
			\right)  x_{\nu} \dot{x}^{\rho}
			+
			\partial_{\nu}			
			\left( 
			\frac{\partial_{\omega} \bar{\phi} \partial_{\sigma} \phi - 
			\partial_{\omega}\phi \partial_{\sigma} \bar{\phi}}
			{1+ \vert \phi \vert^{2}}
			\right)	
		\right)
		\psi^{\omega} \psi^{\sigma}
	\right.
\right.
\nonumber
\\
& -
\left.
	\left.
    \frac{q}{2}\chi 
		\left(
		\partial_{\rho} A_{\omega} \dot{x}^{\rho} \dot{x}^{\omega}
		\psi_{\nu}
		-\psi_{\sigma}
		\frac{q \sqrt{a}}{2 \pi i} 		
		\partial_{\nu} 
		\left(
		\frac{\bar{\phi} \partial^{\sigma} \phi - \phi \partial^{\sigma} 				\bar{\phi}}{1+ \vert \phi \vert^{2}}
		\right)
		\right)
	\right]
\right\}
\nonumber
\\
&
= \dot{x}^{\mu} \frac{d}{d \tau} 
\left\{ 
\ln 
	\left[
	\frac{1}{2e} \dot{x}^{2} + \frac{1}{2} e m^2 + 
	\frac{q \sqrt{a}}{4 \pi i} 
		\left(
		\frac{\bar{\phi} \partial_{\rho} \phi - \phi \partial_{\rho} 				\bar{\phi}}{1+ \vert \phi \vert^{2}}
		\right)
		\dot{x}^{\rho}
	\right.
\right.
\nonumber
\\
& +
\left.
	\left.
	\frac{i}{2}
	\left(
	\dot{\psi}_{\lambda} \psi^{\lambda} - \dot{\psi}_{5} \psi^{5}
	\right) 
	- \frac{i}{2} \chi \theta_{\lambda} \dot{x}^{\lambda}
	+i n \chi \theta_5 
	+  
	\frac{q^2 \sqrt{a}}{4\pi i} e
	\left(
	\frac{\partial_{\rho} \bar{\phi} \partial_{\sigma} \phi - 
	\partial_{\rho}\phi \partial_{\sigma} \bar{\phi}}
	{1+ \vert \phi \vert^{2}}
	\right)
	S^{\rho \sigma}		
	\right]	
\right\}
\, .
\label{geodesic-equation-Finsler-spinning}
\end{align}

\section*{Conflict of interest statement}

On behalf of all authors, the corresponding author states that there is no conflict of interest.


\end{document}